\documentclass[10pt]{article}
\begin{document}
\centerline {\large The effect of noncommutativity of the conservation laws on} 
\centerline {\large the development of thermodynamical and gas dynamical instability}
\centerline  {\large  (The method of skew-symmetric differential forms)}
\centerline {\it L.~I. Petrova }
\centerline{\it Moscow State University, Russia, e-mail: ptr@cs.msu.su}
\bigskip

In the works by the author[1,2] it has been shown that the conservation laws 
for material media (the conservation laws for energy, linear momentum, angular 
momentum, and mass, that establish a balance between the variation of a 
physical quantity and the corresponding external action), turn out to be 
noncommutative. The noncommutativity of the conservation laws that leads to 
an emergence of internal forces and an appearance of the nonequilibrium is 
a cause of development of instability in  material media (material systems). 

These results were obtained with the help of the mathematical apparatus of 
skew-symmetric differential forms. To study the causes of developing 
instability it is necessary to inspect the evolutionary relation obtained 
from equations of the balance conservation laws and to analyze the differential 
form commutator that enters into this relation. 

In the present work the method of skew-symmetric differential forms has been 
applied in study of the thermodynamical and gas dynamical systems. 

It was shown that the principle of thermodynamics follows from two conservation 
laws, namely, the balance conservation laws for energy and linear momentum. In 
this case the second principle of thermodynamics, from which the state function 
(which specifies a state of the thermodynamical system) is obtained, follows 
from the first principle. 

A mechanism of development of instability in gas dynamical systems is described 
and there are explained such processes as emergence of waves, vortices, 
turbulent pulsations and so on. 

\section {Analysis of principles of thermodynamics}

The thermodynamics is based on the first and second principles 
of thermodynamics that were introduced as postulates [3].

Let us show that the first
principle of thermodynamics follows from the balance conservation laws for
energy and linear momentum and is valid for the case when the heat influx is
the only external action. The mathematical representation of the first 
principle of thermodynamics appears to be a nonidentical relation, and this 
points out that the balance conservation laws are noncommutative and the 
state of thermodynamic system is nonequilibrium. 

The second principle of
thermodynamics with the equality sign is obtained from the first principle
under realization of the integrability condition when a role of the
integrating factor plays the inverse temperature, and this corresponds to
a locally equilibrium state of the
system when the temperature appeared to be realized. The second principle of
thermodynamics with the inequality sign takes into account an availability
of some actions other than the heat influx.

As it is well known, the first principle of thermodynamics can be presented
in the form
$$dE+\delta w\,=\,\delta Q $$
where $dE$ is the change of energy of the thermodynamic system, $\delta w$ is 
the work done by the system (this means that $\delta w$ is expressed in terms 
of the system parameters), $\delta Q$ is an amount of the heat putted into the
system (i.e. the external action onto the system). Since the term $\delta w$
can be expressed in terms of the system parameters and specifies a real
(rather than virtual) change, it can be designated by $dw$, and hence,
the first principle of thermodynamics takes the form
$$dE\,+\,dw\,=\,\delta Q\eqno(1.1)$$

What is the difference between the first principle of thermodynamics and
the balance conservation laws?

For thermodynamic system the balance conservation law of energy can be written 
as 
$$ dE\,=\,\delta Q\,+\,\delta G \eqno(1.2)$$
where by $\delta G $ we designate energetic actions with the exception
of heat influx. For thermodynamic system the balance conservation law for
linear momentum (a change of linear momentum of the system in its
dependence on the force mechanical action onto the system) can be written as
$$dw\,=\,\delta W \eqno(1.3)$$
Here $\delta W$ stands for the force (mechanical) action onto the system (for
example, an external compression of the system, an influence of boundaries
and so on).

If to combine relations (1.2) and (1.3), one can obtain the relation
$$dE\,+\,dw\,=\,\delta Q\,+\,\delta G\,+\,\delta W \eqno(1.4)$$

By comparing relation (1.4) that follows from the balance
conservation laws for energy and linear momentum and relation (1.1),
one can see that they coincide if the heat influx is the only external
action onto the thermodynamic system ($\delta W\,=\,0$ and $\delta G \,=\,0$).

Thus, the first principle of thermodynamics follows from the balance
conservation laws for energy and linear momentum. (This is analogous to
the evolutionary relation for the thermodynamic system [1]). 

The significance of the first principle of thermodynamics, (as well as that
of the evolutionary equation), consists in that it clarifies a nature of
interactions of two balance conservation laws (rather than it only
corresponds to the conservation law for energy).

Since $\delta Q$ is not a differential (closed form),
relation (1.1) which corresponds to the first principle of thermodynamics, as 
well as the evolutionary relation,  appears to be a nonidentical nonintegrable
relation. This  points to a noncommutativity of the balance conservation
laws (for energy and linear momentum) and to a nonequilibrium state of the
thermodynamic system.
Although the form $dE\,+\,p\,dV$  consists of differentials, in the general
case without the integrating factor it is not a differential 
(closed exterior differential forms) because of
that its terms depend on different variables, namely, the first term is
determined by variables that specifies the internal structure of elements,
and the second term depends on variables that specify an interaction
between elements, for example, the pressure. The commutator of the form 
$dE\,+\,pdV$ is nonzero. This points to an availability of intrnal forces 
and the nonequilibrium state of the thermodynamical system.

As it follows from the analysis of the evolutionary relation [1], the 
transition to a locally equilibrium state must correspond to
realization of the additional condition (i.e. the integrability condition).
If this condition be satisfied, from the nonidentical evolutionary
relation, which corresponds to the first principle of thermodynamics, it
follows an identical relation. It is an identical relation
that corresponds to the second principle of thermodynamics.

Let us consider the case when the work performed by the system is carried out
through the compression. Then $dw\,=\,p\,dV$ (here $p$ is the pressure and $V$
is the volume) and $dE\,+\,dw\,=\,dE\,+\,p\,dV$. As it is known, the form
$dE\,+p\,dV$ can become a differential if there is the integrating factor 
$\theta$ (a quantity which depends only on the characteristics of the system),
where $1/\theta\,=\,pV/R$ is called the temperature $T$ [3]. 
In this case the form $(dE\,+\,p\,dV)/T$ turns out to be a differential 
(interior) of some quantity that referred to as entropy $S$:
$$(dE\,+\,p\,dV)/T\,=\,dS \eqno(1.5)$$

If the integrating factor $\theta=1/T$ has been
realized, that is, relation (1.5) proves to be satisfied, from relation (1.1), 
which corresponds to the first principle of thermodynamics, 
it follows
$$dS\,=\,\delta Q/T \eqno(1.6)$$
This is just the second principle of thermodynamics for reversible processes.
It takes place when the heat influx is the only action onto the system.
If in addition to the heat influx the system experiences a certain mechanical
action $\delta W$
(for example, an influence of boundaries), then according to relation (1.4)
from relation (1.5) we obtain
$$dS\,=\,(dE+p\,dV)/T\,=\,(\delta Q+\delta W+\delta G)/T \eqno(1.7)$$
from which it follows
$$dS\, >\,\delta Q/T \eqno (1.8)$$
that corresponds to the second principle of thermodynamics for irreversible
processes.

Relations (1.6), (1.8) that can be written as
$$dS\,\geq \,\delta Q/T, \eqno(1.9)$$
express the second principle of thermodynamics. (It is well to bear in mind
that the differentials in relations (1.5), (1.6), (1.8), (1.9)  are
not total differentials.
They are satisfied only with the presence of the integrating factor, namely,
the temperature, which depends on the system parameters).

Thus, the first principle of thermodynamics follows from the balance 
conservation laws for energy and linear momentum, and the second principle of 
thermodynamics follows from the first one. The second principle of 
thermodynamics with the equality sign follows from the first principle under 
the fulfillment of the condition of integrability, i.e. a realization of the 
integrating factor (the inverse temperature). (This corresponds to the 
transition from a nonequilibrium state to a locally equilibrium state. 
Phase transitions, the origin of fluctuations, etc are examples of such 
transitions). In this case entropy proves to be the state function.
And the second principle of thermodynamics with the
inequality sign takes into account an availability in actual processes other
actions besides the heat influx. For this case entropy is a functional.

In the case examined above a differential of entropy (rather than entropy 
itself) becomes a closed form. $\{$In this case entropy  manifests itself 
as the thermodynamic potential, namely, the function of state. To the
pseudostructure there corresponds
the state equation that determines the temperature dependence on the
thermodynamic variables$\}$.

For entropy to be a closed form itself (a conservative quantity), one more
condition must be realized. Such a condition could be the realization of the
integrating direction, an example of that is the speed of sound:
$a^2\,=\,\partial p/\partial \rho\,=\,\gamma\,p/\rho$. In this case it is valid
the equality $ds\,=\,d(p/\rho ^{\lambda })\,=\,0$ from which it follows that
entropy $s\,=\,p/\rho ^{\lambda }\,=\hbox{const}$ is a closed form (of zero 
degree). $\{$However it does not mean that a state of the gaseous system is 
identically isoentropic. Entropy is constant only along the integrating 
direction (for example, on the adiabatic curve or on the front of the sound 
wave), whereas in the direction normal to the integrating direction the 
normal derivative of entropy has a break$\}$.

Under realization of the integrating direction the transition from the 
variables $E$, $V$ to the variables $p$, $\rho$ is a degenerate transform.

It worth underline that both temperature and the speed of sound are not
continuous thermodynamic variables. They are variables that are realized
in the thermodynamic processes if the thermodynamic system has any degrees of
freedom. One can see the analogy between the inverse temperature and the speed 
of sound: the inverse temperature is the integrating factor and the speed of 
sound is the integrating direction.
$\{$Notice that in actual processes a total state of the thermodynamic 
system is nonequilibrium one and the commutator of the form $dE\,+\,pdV$ is 
nonzero. A quantity that is described by a commutator and acts as an 
internal force can grow. Prigogine [4] defined this as the "production
of the excess entropy". It is the increase of the internal force that is
perceived as the growth of entropy in the irreversible processes$\}.$

A closed static system, if left to its own devices, can tend to a state
of total thermodynamic equilibrium. This corresponds to tending the system
functional to its asymptotic maximum. In the dynamical system the tending
of the system to a state of total thermodynamic equilibrium can be violated
by dynamical processes and transitions to a state of local equilibrium.

\section {The development of the gas dynamic instability}

To study a development of instability it is necessary to analyze the balance
conservation laws.

For example, we take the simplest gas dynamical system, namely, a flow
of  ideal (inviscous, heat nonconductive) gas [5].

Assume that gas is a thermodynamic system in the state of local equilibrium
(whenever the gas dynamic system itself may be in nonequilibrium
state), that is, it is satisfied the relation [3]
$$Tds\,=\,de\,+\,pdV \eqno(2.1)$$
where $T$, $p$ and $V$ are the temperature, the pressure and the gas 
volume, $s$, $e$ are entropy and internal energy per unit volume.

Let us introduce two frames of reference: an inertial one that is not connected 
with the material system and an accompanying frame of reference that is 
connected with the manifold formed by the trajectories of the material system 
elements. (Both Euler's and Lagrange's systems of coordinates
can be examples of such frames).

In the inertial frame of reference the Euler equations are the balance
conservation laws for energy, linear momentum and mass of ideal gas [5].

The equation of the balance conservation law of energy for ideal gas can
be written as
$${{Dh}\over {Dt}}- {1\over {\rho }}{{Dp}\over {Dt}}\,=\,0 \eqno(2.2)$$
where $D/Dt$ is the total derivative with respect to time (if to designate
the spatial coordinates by $x_i$ and the velocity components by $u_i$,
$D/Dt\,=\,\partial /\partial t+u_i\partial /\partial x_i$). Here  $\rho=1/V $
and $h$ are respectively the mass and the entalpy densities of the gas.

Expressing entalpy in terms of internal energy $e$ with the help of formula
$h\,=\,e\,+\,p/\rho $ and using relation (2.1) the balance conservation law
equation can be put to the form
$${{Ds}\over {Dt}}\,=\,0 \eqno(2.3)$$

And respectively, the equation of the balance conservation law for linear
momentum can be presented as [5,6]
$$\hbox {grad} \,s\,=\,(\hbox {grad} \,h_0\,+\,{\bf U}\times \hbox {rot} {\bf U}\,-{\bf F}\,+\,
\partial {\bf U}/\partial t)/T \eqno(2.4)$$
where ${\bf U}$ is the velocity of the gas particle, 
$h_0=({\bf U \cdot U})/2+h$, ${\bf F}$ is the mass force. The operator $grad$ 
in this equation is defined only in the plane normal to the trajectory.
[Here it was tolerated a certain incorrectness. Equations (2.3), (2.4) are
written in different forms. This is connected with difficulties when
deriving these equations themselves.
However, this incorrectness will not effect on results of the qualitative
analysis of the evolutionary relation obtained from these equations.]

Since the total derivative with respect to time is that along the trajectory,
in the accompanying frame of reference equations (2.3) and (2.4)
take the form:
$${{\partial s}\over {\partial \xi ^1}}\,=\,0 \eqno (2.5)$$
$${{\partial s}\over {\partial \xi ^{\nu}}}\,=\,A_{\nu },\quad \nu=2, ... \eqno(2.6)$$
where $\xi ^1$ is the coordinate along the trajectory,
$\partial s/\partial \xi ^{\nu }$
is the left-hand side of equation (2.4), and $A_{\nu }$ is obtained from the
right-hand side of relation (2.4). 

\{In the common case when gas is
nonideal equation (2.3) can be written in the form
$${{\partial s}\over {\partial \xi ^1}} \,=\,A_1 \eqno (2.7)$$
where $A_1$ is an expression that depends on the energetic actions. In the case 
of ideal gas $A_1\,=\,0$ and equation (2.7) transforms into (2.5). In the case
of the viscous heat-conductive gas described by a set of the Navier-Stokes 
equations, in the inertial frame of reference the expression $A_1$ can be 
written as [5]
$$A_1\,=\,{1\over {\rho }}{{\partial }\over {\partial x_i}}
\left (-{{q_i}\over T}\right )\,-\,{{q_i}\over {\rho T}}\,{{\partial T}\over {\partial x_i}}
\,+{{\tau _{ki}}\over {\rho }}\,{{\partial u_i}\over {\partial x_k}} \eqno(2.8)$$
Here $q_i$ is the heat flux, $\tau _{ki}$ is the viscous stress tensor.
In the case of reacting gas extra terms connected with the chemical
nonequilibrium are added [5].\}

Equations (2.5) and (2.6) can be convoluted into the equation
$$ds\,=\,\omega \eqno(2.9)$$
where $\omega\,=\,A_{\mu} d\xi ^{\mu}$ is the first degree differential form
(here $\mu =1,\,\nu $).

Relation (2.9) is an evolutionary relation for gas dynamic system
(in the case of local thermodynamic equilibrium). Here $\psi\,=\,s$.
\{It worth notice that in the evolutionary relation for thermodynamic
system the dependence of entropy on thermodynamic variables is investigated
(see relation (2.1)), whereas in the evolutionary relation for gas dynamic
system the entropy dependence on the space-time variables is considered\}.

If relation (2.9) appears to be identical one (if the form $\omega $
be the closed form, and hence it is a differential),
one can obtain a differential of entropy $s$ and find entropy as a function
of space-time coordinates. It is entropy that will be the gas dynamic function
of state. \{It should underline once again that entropy as
a thermodynamic function of state is not gas dynamic function of state\}.
The availability  of the gas dynamic function of state would point to the
equilibrium state of the gas dynamic system.
If relation (2.9) be not identical (this takes place for actual processes),
then from this relation the differential of entropy $s$ cannot be defined.
This will point to an absence of the gas dynamic function of state and
nonequilibrium state of the system. Such nonequilibrium is a cause of the
development of instability.

Since the nonequilibrium is produced by internal forces that are described
by the commutator of the form $\omega $, it becomes evident that a cause
of the gas dynamic instability is something that contributes into the
commutator of the form $\omega $. Without accounting for terms that
are connected with a deformation of the manifold formed by the trajectories
the commutator can be written as
$$K_{1\nu }\,=\,{{\partial A_{\nu }}\over {\partial \xi ^1}}\,-\,{{\partial A_1}\over
{\partial \xi ^{\nu }}} \eqno(2.10)$$

From the analysis of the expression $A_{\nu }$ and with taking into account
that $A_1\,=\,0$ one can see that terms
that are related to the multiple connectedness of the flow domain (the second
term of  equation (2.4)), the nonpotentiality of the external forces
(the third term in (2.4)) and the nonstationarity of the flow (the forth term
in (2.4)) contribute into the commutator. \{In the general case the terms 
connected with transport phenomena and physical and chemical processes will 
contribute into the commutator (see equation (2.8) \}.

One can see that the development of instability is caused by
not a simply connectedness of the flow domain,  nonpotential  external
(for each local domain of the gas dynamic system) forces, a nonstationarity
of the flow, transport phenomena. (In common case 
on the gas dynamic instability it will effect the thermodynamic,
chemical, oscillatory, rotational, translational nonequilibrium).

All these factors lead to emergence of internal forces, 
that is, to nonequilibrium and to development
of various types of instability. (It can be noted that for the case of
ideal gas Lagrange derived a condition of the eddy-free stable flow.
This condition is as follows: the domain must be simple connected one,
forces must be potential, the flow
must be stationary. One can see, that  under fulfillment of these conditions
there are no terms that contribute into the commutator).

And yet for every
type of instability one can find an appropriate term giving contribution
into the evolutionary form commutator, which is responsible for this type
of instability.
Thus, there is an unambiguous connection between the type of instability
and the terms that contribute into  the evolutionary form commutator in the
evolutionary relation. \{In the general case one has to consider the
evolutionary relations that correspond to the balance conservation laws 
for angular momentum and mass as well\}.

Whether the gas dynamic system can get rid of
the internal force and transfer into the locally equilibrium state?
(The mechanism of evolutionary processes in material media and the transition 
from the nonequilibrium state to the locally equilibrium state is presented 
in detali in works [1,2]).

The locally equilibrium state corresponds the state differential that is a
closed form. 
The transition from evolutionary differential form $\omega$ to closed form, 
that would correspond to the transition
from the nonequilibrium state of the system to the locally equilibrium state,
is possible only as {\it the degenerate transform}, i.e. the
transform that does not conserve the differential.
The evolutionary differential form $\omega$,
involved into evolutionary relation, is an unclosed one for real processes. The
commutator, and hence the differential, of this form is nonzero.
The locally equilibrium state corresponds the state differential that is a
closed form. The differential of the closed form is zero. 

To the degenerate transform it must correspond a vanishing of some functional 
expressions. 
Such functional expressions may be Jacobians, determinants, the Poisson
brackets, residues and others. It is obvious that the condition of degenerate 
transform has to be due to the gas dynamic system properties. This may be, for
example, the availability of any degrees of freedom in the gas dynamic system. 

If the transformation is degenerate, from the unclosed evolutionary form it 
can be obtained a differential form closed on some structure (pseudostructure).  
The differential of this form equals zero. That is, it is 
realized the transition 

$d\omega\ne 0 \to $ (degenerate transform) $\to d_\pi \omega=0$, 
$d_\pi{}^*\omega=0$  

Here ${}^*\omega$ is the dual form (the equation of the pseudostructure 
$\pi$ [1]). 

(The degenerate transformation is realized as a transition from the
accompanying noninertial coordinate system to the locally inertial that).

On the pseudostructure $\pi$ evolutionary relation (2.9) transforms into
the relation
$$
d_\pi s=\omega_\pi\eqno(2.10)
$$
which proves to be the identical relation. Indeed, since the form
$\omega_\pi$ is a closed one, on the pseudostructure it turns
out to be a differential of some differential form. 

The identical relation (2.10) obtained from the nonidentical evolutionary 
relation under degenerate transform integrates the state differential and the 
closed (inexact) exterior differential form [1]. The availability of the state 
differential indicates that the material system state becomes a locally 
equilibrium state (that is, the local domain of the system under consideration 
changes into the equilibrium state). The availability of the exterior closed 
on the pseudostructure differential form means that the physical structure is 
present. This shows that the transition of material system into the locally 
equilibrium state is accompanied by the origination of physical structures. 

As it has been shown in works [1,2], in the material system  origination of 
a physical structure reveals as a new
measurable and observable formation that spontaneously arises in the material 
system. In the physical process this formation is 
spontaneously extracted from the local domain of the material system and so 
it allows the local domain of material system to get rid of an internal 
force and come into a locally equilibrium state.

The gas dynamic formations that correspond to these physical structures are
shocks, shock waves, turbulent pulsations and so on. 

It is evident that the characteristics of the formation (intensity, vorticity, 
absolute and relative speeds of propagation of the formation) are determined 
by the evolutionary form and its commutator and by the material system characteristics.

In works [1,2] it has been shown a connection of characteristics of 
the formation orignated with characteristics of the evolutionary forms, the 
evolutionary form commutators obtained from closed forms, and the 
material system characteristics. In the present work we shall not fix 
our attention on these problems.

Additional degrees of 
freedom are realized as the condition of the degenerate transform, namely, 
vanishing of determinants, Jacobians of transforms, etc. These conditions 
specify the integral surfaces (pseudostructures):
the characteristics (the determinant of coefficients at the normal derivatives
vanishes), the singular points (Jacobian is equal to zero), the envelopes
of characteristics of the Euler equations and so on. Under passing
throughout the integral surfaces
the gas dynamic functions or their derivatives suffer breaks. Below we present
the expressions for calculation of such breakes of derivatives in the direction
normal to characteristics (and to trajectories). 

The extraction of some formation from the local domain of the system is 
accompanied by emergence of the break surfaces (contact breaks). These 
breaks are connected with trajectories of the material system elements.

Let as analyze which types of instability and what gas dynamic  formation 
can originate under given external action.

1). {\it Shock, break of diaphragm and others}. The instability originates
because
of nonstationarity. The last term in equation (2.4) gives a contribution
into the commutator. In the case of ideal gas whose flow is described by
equations of the hyperbolic type the transition to the locally equilibrium
state is possible on the characteristics and their envelopes. The
corresponding structures are weak shocks and shock waves.

2).{\it Flow of ideal (inviscous, heat nonconductive) gas around bodies
Action of nonpotential forces}. The instability develops because of
the multiple connectedness of the flow domain and a nonpotentiality of the
body forces. The contribution into the commutator comes from the second and
third terms of the right-hand side of  equation (2.4). Since the gas is ideal 
one and $\partial s/\partial \xi ^1=A_1=0$, that is, there is no contribution
into the each fluid particle, an instability of convective type develops.
For $U>a$ ($U$ is the velocity of the gas particle, $a$ is the speed of sound)
a set of equations of the balance conservation laws belongs to the
hyperbolic type and hence the transition to the locally equilibrium state is
possible on the characteristics and on the envelopes of characteristics
as well, and weak shocks and shock waves are the structures of the system.
If $U<a$ when the equations are of elliptic type, such a transition is
possible only at singular points. The structures emerged due to a convection 
are of the vortex type. Under long acting the large-scale structures
can be produced.

3. {\it Boundary layer}. The instability originates due to the multiple
connectness of the domain and the transport phenomena (an effect of
viscosity and thermal conductivity). Contributions into the commutator produce
the second term in the right-hand side of equation (2.4) and the second and
third terms in expression (2.8). The transition to the locally equilibrium
state is allowed at singular points. because in this case
$\partial s/\partial \xi^1=A_1\neq 0$, that is, the external exposure acts
onto the gas particle separately, the development of instability and the
transitions to the locally equilibrium state are allowed only in
an individual fluid particle. Hence, the structures emerged behave as
pulsations. These are the turbulent pulsations.

$\{$It is commonly believed that the instability is an emergence of any
structures in the gas dynamic flow. From this viewpoint the laminar boundary
layer is regarded as stable one, whereas the turbulent layer regarded as
unstable layer. However the laminar boundary layer cannot be regarded as
a stable one because of the fact that due to the not simple connectedness
of the flow domain and the transport processes the
instability already develops although any structures do not originate.
In the turbulent boundary layer the emergence of pulsations is the transition
to the locally equilibrium state, and the pulsations themselves are local
formations. The other matter, due to the global nonequilibrium the locally
equilibrium state is broken up and the pulsations weaken$\}$.

Studying the instability on the basis of the analysis of entropy
behavior was carried out in the works by Prigogine and co-authors [7].
In that works entropy was considered as the thermodynamic function
of state (though its behavior along the trajectory was analyzed).
By means of such state function one can trace the development (in gas
fluxes) of the hydrodynamic instability only. To investigate the gas
dynamic instability it is necessary to consider entropy as the gas dynamic
state function, i.e. as a function of the space-time coordinates.
Whereas for studying the thermodynamic instability one has to analyze
the commutator constructed by the mixed derivatives of entropy with respect
to the thermodynamic variables, for studying the gas dynamic instability
it is necessary to analyze the commutators
constructed by the mixed derivatives of entropy with respect to the space-time
coordinates.

In conclusion it should be said a little about modelling instable flows.
As it is known, some authors tried to account for the development of 
instability by means of improving the equations modelling the balance 
conservation laws (for example, 
by introducing the high-order moments) or by introducing additional
equations. However, such attempts give no satisfactory results. To describe
the nonequilibrium flow and the emergence of the gas dynamic structures
(waves, vortices, turbulent pulsations) one must add
the evolutionary relation obtained from the balance conservation law equations
to the balance conservation law equations.  Under numerical modelling
the gas flows one has to trace for the transition from the evolutionary
nonidentical relation to the identical relation (for the transition from
an evolutionary unclosed form to an exterior closed form), and this will
point to the emergence of a certain physical structure.

\bigskip
Below we present an example of calculating the breaks of derivatives of the
gas dynamic functions that are necessary for numerical analysis of gas
dynamic flows.

\subsection*{Breaks of normal derivatives on characteristics and trajectories}

\bigskip
While studying the effects connected with the origination of the vorticity one 
can notice a certain specifics of numerical solving the Euler equations [5].
This may be demonstrated by the following example. Assume, that the initial
conditions correspond to the isoentropic flow, that is, entropy is the same
along all trajectories. For ideal gas under consideration entropy conserves
along the trajectory. From this it follows that entropy has to conserve
during all time of flow. However, in actual cases (unsteady flow, flow along
a body, heterogeneous medium) the derivative of entropy along the direction
normal to trajectory suffers the break. Thus we have that, from one side,
entropy (function) must be constant and, from other hand, its derivatives
suffer the breaks. This contradiction is resolved with taking into account
the fact that the break of derivative is compensated by changing the stream
function or bending the trajectory. It is this effect that must be accounted 
for in the process of numerical calculation.  In particular, when calculation
the one-dimensional nonstationary nonisoentropic flow of gas the conditions
on the characteristics includes the derivative of entropy with respect to the
coordinate normal to trajectory (in space of two variables, namely, time and
coordinate). To calculate this derivative one must know the break 
of derivative of entropy. This can be obtained from the relations
that connects the breaks of derivatives of the gas dynamic functions.

These relations are found from the dynamic conditions of a consistency of
the Euler equations. In paper [8] the dynamical conditions of consistency
of the Euler equations for the case $p=f(\rho)$ were considered.

In the present work in a similar manner it is analyzed the case when
$p=f(\rho ,s)$, where $s$ is the entropy, and the relations that connect
the breaks of
derivatives of the functions describing the particle velocity, the sound speed,
and entropy are obtained. These relations enable one to carry out
numerical calculations of the nonisoentropic gas flows.

The scheme of obtaining these relations for one-dimensional nonstationary
equations is the following. At the beginning, the Euler equations are written
down. Then the equations for characteristics and the conditions on
characteristics are derived. Kinematical conditions of consistency [8]
that mutually
connect the breaks of derivatives of the gas dynamic functions are
written down. These
conditions are substituted into the Euler equations. As a result, a
homogeneous set of equations for the breaks of derivatives of the
functions desired is obtained. On the characteristical surface the
determinant of this set equals zero, and from this it is found the
nontrivial solution for the breaks of derivatives of the functions desired
in their dependence on a value of one of others.

If to take $U$ (the gas velocity), $a$(the sound speed), and $s$,
the following relations are obtained:

1) In the direction normal to the trajectory
the derivatives of the sound speed and entropy suffer breaks (the derivative
of velocity does not suffer a break). These breaks are connected between
them by the relation:
$$\left [{{\partial a}\over {\partial \eta _1}}\right ]=\left [{{\partial s}\over
{\partial \eta _1}}\right ] {a\over {2\gamma s}}$$
where $\eta _1$ is the direction normal to the trajectory, $\gamma$
is the Poisson constant.

2) In the direction normal to the characteristics the derivatives of the
gas velocity and the speed of sound suffer breaks (the derivative of
entropy does not suffer break). These breaks are connected between them
by the relation:
$$\left [{{\partial u}\over {\partial \eta _{+-}}}\right ]= \pm \left [{{\partial a}
\over {\partial \eta _{+-}}}\right ]{2\over{\gamma -1}}$$
where $\eta _{+-}$ are the directions normal to the corresponding
characteristics.

1. Petrova L.~I., Exterior and evolutionary skew-symmetric differential forms 
and their role in mathematical physics. http://arXiv.org/pdf/math-ph/0310050

2. Petrova L.~I., Conservation laws. Their role in evolutionary processes.   
(The method of skew-symmetric differential forms). 
http://arXiv.org/pdf/math-ph/0311008

3. Haywood R.~W., Equilibrium Thermodynamics. Wiley Inc. 1980.  

4. Prigogine I., Introduction to Thermodynamics of Irreversible 
Processes. --C.Thomas, Springfild, 1955.  

5. Clark J.~F., Machesney ~M., The Dynamics of Real Gases. Butterworths, 
London, 1964. 

6. Liepman H.~W., Roshko ~A., Elements of Gas Dynamics. Jonn Wiley, 
New York, 1957.  

7. Glansdorff P., Prigogine I. Thermodynamic Theory of Structure, Stability 
and Fluctuations. Wiley, N.Y., 1971.  

8. Smirnov V.~I., A course of higher mathematics. -Moscow, 
Tech.~Theor.~Lit. 1957, V.~4 (in Russian). 

\end{document}